\documentclass{iopart}

\usepackage{color}
\usepackage{graphicx}
\usepackage{latexsym}
\usepackage{bm}

\begin{document}

\title[ASEP in 1D chains with long-range links]{Asymmetric simple exclusion 
process in one-dimensional chains with long-range links}

\author{Mina Kim$^1$, Ludger Santen$^2$ and Jae Dong Noh$^{1,3}$}

\address{$^1$ Department of Physics, University of Seoul, Seoul 130-743, Korea}
\address{$^2$ Theoretische Physik, Universit\"at des Saarlandes,  
66041 Saarbr\"ucken, Germany} 
\address{$^3$ School of Physics, Korea Institute for Advanced Study, Seoul
130-722, Korea}
\ead{jdnoh@uos.ac.kr}

\date{\today}
\begin{abstract}
We study the boundary-driven asymmetric simple exclusion process~(ASEP) 
in a one-dimensional chain with long-range links. 
Shortcuts are added to a chain by connecting $pL$ different pairs of sites 
selected randomly where $L$ and $p$ denote the chain length and 
the shortcut density, respectively. 
Particles flow into a chain at one boundary at rate $\alpha$ 
and out of a chain at the other boundary at rate $\beta$, while 
they hop inside a chain via nearest-neighbor bonds and long-range shortcuts. 
Without shortcuts, the model reduces to the boundary-driven ASEP in a 
one-dimensional chain which displays the low density, high density, 
and maximal current phases. 
Shortcuts lead to a drastic change. Numerical simulation studies suggest 
that there emerge three phases; an empty phase with $ \rho = 0 $, 
a jammed phase with $ \rho = 1 $, and a shock phase with $ 0<\rho<1$ 
where $\rho$ is the mean particle density. 
The shock phase is characterized with a phase separation between an empty 
region and a jammed region with a localized shock between them. 
The mechanism for the shock formation and the non-equilibrium phase 
transition are explained by an analytic theory based on a mean-field 
approximation and an annealed approximation.
\end{abstract}
\maketitle

\section{Introduction}\label{sec1}
The asymmetric simple exclusion process~(ASEP) has been widely studied 
 in the past decades~\cite{Derrida98}.
It is a nonequilibrium driven diffusive system of particles subject to 
the exclusion interaction. 
Despite its simplicity, the ASEP 
 describes various nonequilibrium processes such as 
bio-polymerization~\cite{MacDonald68}, surface 
growth~\cite{Spohn06}, traffic flow~\cite{Nagel92,Chowdhury00}, 
for example.
Furthermore, the ASEP in one dimension is exactly solvable via the 
Bethe ansatz~\cite{Gwa92} and the matrix product 
ansatz~\cite{Derrida93} or direct solution of recursion 
relations~\cite{Schutz93}.
The exact solution contributes to deeper understanding of fluctuation 
phenomena~\cite{Derrida98_1,Lee99} and nonequilibrium phase
transitions~\cite{Schutz93}.

Most studies on the ASEP have been performed on one-dimensional~(1D) chains. 
On the other hand, there are many quasi-1D systems 
which involve long-range links. If a polymer chain folds randomly, there 
arise contacts between different polymer segments which are far apart along a 
backbone~\cite{deGennes}. Those contacts can be regarded as long-range links 
for a transport process on a network of polymer chains. 
A gene regulatory protein
diffuses along the DNA chain to search for its target gene~\cite{Berg81}. 
It can make a long-range jump by dissociation from and reassociation with 
the DNA. Recent studies show that real world traffic networks have a complex
structure with long-range links~\cite{Watts98,BA}. 
In this respect, interests are growing in the study of the ASEP on complex
networks~\cite{Szavits-Nossan,Ha07,Otwinowski09}.

In this work, we address the question: What is the transport capacity of 
a complex network for particles interacting via mutual exclusion? 
In order to contribute to this issue we consider the ASEP on a 1D chain with 
long-range links, called shortcuts, for open boundary conditions. 
It will turn out that the shortcuts cause a drastic change in the phase 
diagram of the ASEP.

We start with a brief review of the 1D ASEP. The original model is defined on a 
1D lattice. The lattice sites are either occupied by at most one particle or empty. 
Multiple occupancy on a site is prohibited~[exclusion interaction].
Particles may hop to the left and right with a bias to one direction~[ASEP].
A closely related equilibrium process is the so-called symmetric simple 
exclusion process~(SSEP) where particles hop in both directions with equal 
rates.
For periodic boundary conditions, the model has a trivial steady state 
where every microscopic configuration is equally likely irrespective 
of the hopping bias~\cite{Gwa92}. However, the bias is relevant for the 
dynamic scaling 
behavior. In the context of growing interfaces, the SSEP belongs to 
the Edward-Wilkinson~(EW) universality class~\cite{Edwards82}, and
the ASEP to the Kardar-Parisi-Zhang~(KPZ) universality
class~\cite{Kardar86}. 
Both classes are characterized by the power-law scaling 
$\tau \sim \xi^z$ between characteristic time and length scales. 
The scaling exponent is given by $z=2$ for the EW class and $z=3/2$ 
for the KPZ class~\cite{Gwa92,Kim95}.

For open boundary conditions the chain is coupled to 
particle reservoirs at both ends. One acts as a particle 
source emitting particles at a rate $\alpha$, and the other as a
sink absorbing particles at a rate $\beta$. 
Interestingly, the system driven by the open boundaries displays 
nonequilibrium phase transitions between low-density~(LD), high-density~(HD), 
and maximal-current~(MC) phases~\cite{Derrida93,Schutz93}. 
The system belongs to the LD phase 
if the capacity of the particle source $\alpha$ controls the particle flux, 
i.e. if $\alpha < \beta$ and $\alpha < p_h/2$, where $p_h$ denotes the hopping 
rate of the particles.
The system belongs to the HD phase when the outgoing rate $\beta$ is smaller 
than $\alpha$ and $p_h/2$. 
The overall particle density is determined by the outgoing rate
$\beta$ and there is a macroscopic congestion of particles. 
When both $\alpha$ and $\beta$ are larger  than $p_h/2$, 
the capacity of the system is limited by the capacity of the bulk. 
Such a phase is called the MC phase, where the overall particle density 
is independent of $\alpha$ and $\beta$.

In the 1D ASEP, particle hopping is a short-ranged process between 
neighboring sites. There are a few attempts to study the effect of a 
long-range hopping on the phase diagram of the ASEP for open boundary 
conditions.
Szavits-Nossan and Uzelac considered the ASEP  with a 
probabilistic long-range hopping~\cite{Szavits-Nossan}.
A particle can hop to any site at a distance 
$l$ with probability $p_{l} \sim 1/l^{\sigma+1}$. 
They obtained that the phase diagram remains the same as that of 
the 1D ASEP for $\sigma>1$. For $\sigma \leq 1$, however, the system does 
not display any phase transition. Ha {\it et al.} considered a boundary-driven
1D ASEP model where particles may perform a short-range hopping between 
nearest-neighbor sites or a long-range hopping~\cite{Ha07}.
Upon a long-range hopping, a particle jumps to an empty site directly 
behind a next particle in front of it. It was found that the long-range 
hopping introduces an instability towards a so-called empty-road phase.
Otwinowski and Boettcher considered the ASEP on a one-dimensional chain 
decorated with hierarchically organized long-range links~\cite{Otwinowski09}. 
This model displays the LD and HD phases. Besides, depending on the way 
the long-range hopping is implemented, an intermediate phase may also be 
realized.  

In this work, we investigate the role of long-range hoppings in a 
driven system in a generic setting. 
For that purpose, we study the ASEP on a 1D chain with open boundary
conditions and additional long range links 
connecting randomly-selected pairs of sites.
The result is a graph similar to a small-world network~\cite{Watts98}. 
Additionally, we couple this network to two particle reservoirs, a particle 
source connected to one boundary site and sink to another. 
In contrast to the 1D ASEP, particle source and sink now are connected via 
the small-world network. This setup describes a generic scenario 
for directed transport through a complex network with limited capacity.
It is generally believed that long-range links make a system defined on a 
small-world network homogeneous~\cite{Dorogovtsev08}.
In contrast, our study shows that the system develops a localized shock 
which separates a 1D backbone into an empty region near 
the entrance and a fully-occupied region near the exit. 
Such an inhomogeneity is caused by the interplay 
between the boundary driving and the long-range hopping. 
The steady-state position of the shock depends on the particle input 
and output rates, which results in an interesting phase diagram.

This paper is organized as follows:  In Sec.~\ref{sec2}, we introduce
a boundary-driven ASEP in a 1D chain with long-range links.
In Sec.~\ref{sec3}, we present numerical results obtained from Monte Carlo 
simulations. Measuring the overall particle density $\rho$, we
obtain a numerical phase diagram which consists of an empty~(E) 
phase, a jammed~(J) phase, and a shock~(S) phase.
The overall particle density takes the value of 
$\rho=0~(1)$ in the E~(J) phase, while it varies continuously in the S
phase. The shock phase is characterized by a localized shock which separates
a 1D backbone into empty and jammed domains. In order to understand 
the mechanism for the shock formation and the phase transition, we develop
an approximate analytic theory. This is presented in Sec.~\ref{sec4}.
We summarize and conclude the paper in Sec.~\ref{sec5}. 

\section{Boundary-driven ASEP with long-range links}\label{sec2}

\begin{figure}[t]
\includegraphics*[width=\columnwidth]{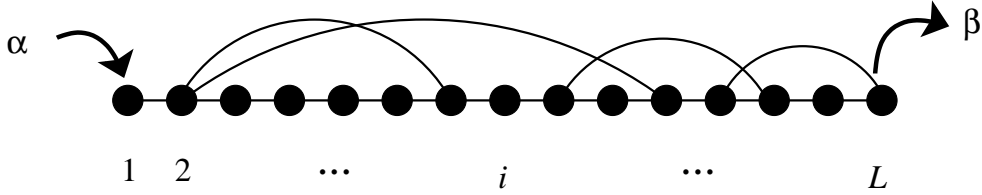}
\caption{A graph consisting of $L$ sites, short-range links, and long-range
links. The site $i=1$ and $L$ are attached to a particle source and sink, 
respectively.}
\label{fig1}
\end{figure}

Consider a graph $\mathcal{G}$ consisting of $L$ sites which are labeled as 
$i=1,2,\cdots,L$. 
Every pair $(i,i+1)$ with $i=1,\cdots,L-1$ is connected with a short-range 
link to form a 1D backbone. In
addition, we select $pL$ pairs of sites at random and add long-range links 
between them~(see Figure~\ref{fig1}). The shortcut density, the total
number of long-range links divided by $L$, is denoted by $p$. Every link has
an orientation: A link between $i$ and $j$ is directed from $i$ to $j$ if
$i<j$ and vice versa. Such a direction will be referred to as a forward
direction. It is assumed that the boundary site $i=1$, called an entrance,
is attached to a particle source 
where particles are fed into the system at  constant rate.
The other boundary site $i=L$, called an exit, is attached
to a particle sink which absorbs particles at a constant rate.

We consider the ASEP process on such a graph. More precisely the particle 
dynamics is defined as follows: 
A particle enters the system at a rate denoted as $\alpha$. 
Particles which have entered the system are then allowed to hop along 
forward links at rate one. 
If a particle is located at a site which is connected to others via multiple 
forward links one of these is randomly chosen. 
For each forward link the selection probability is given by 
$1/k_i$ where $k_i$ denotes the number of forward links at a given site $i$.  
A particle can leave the system at site $L$ (the exit)  at
a rate denoted as $\beta$. Throughout the process, particles are
subject to the exclusion principle which forbids multiple occupancy.
Therefore any trial move violating the exclusion principle is rejected.

One can represent the structure of a graph $\mathcal{G}$
with an adjacency matrix $\bm{A}$ whose elements 
$A_{ij}$ take the value of $1$ 
only if there is a forward link from site $j$ to $i$. 
Due to the link directionality, $A_{ij}=0$ for all $i<j$. 
Then, the hopping probability of a particle from site $i$ to $j$ is given by
\begin{equation}\label{p_hop}
u_{ji} = \frac{A_{ji}}{k_i} \ ,
\end{equation}
where $k_i\equiv \sum_{l>i} A_{li}$ denotes the number of forward
links out of site $i$. 
It should be noticed that the hopping probability is a 
quenched random variable. It varies from one realization of a graph to
another. Consequently, the quenched average over different  realizations of the graph is
necessary.

Here, we are interested in the particle density distribution and the current.
Let $n_i~(=0,1)$ be the occupation number at site $i$. It is a stochastic
variable evolving according to the ASEP dynamics. 
For a given realization of the graph, its average value 
is governed by the time evolution equation 
\begin{equation}\label{dni_dt}
\frac{d}{dt} \langle n_{i} \rangle =
\sum_{j<i} u_{ij} \langle n_j (1-n_i)\rangle - 
\sum_{j>i} u_{ji} \langle n_{i} ( 1-n_j )\rangle
\end{equation}
for $1<i<L$ and
\begin{eqnarray}
\frac{d}{dt}\langle n_1\rangle &=& \alpha(1-\langle n_1\rangle) - 
\sum_{j=2}^L u_{j1} \langle n_1 ( 1-n_j )\rangle \label{dn1_dt} \\
\frac{d}{dt}\langle n_L \rangle &=&
\sum_{j=1}^{L-1} u_{Lj} \langle n_j (1-n_L)\rangle - \beta \langle
n_L\rangle . \label{dnL_dt} 
\end{eqnarray}
Here the angle bracket $\langle(\cdot) \rangle$ represents the average 
of a quantity $(\cdot)$ for different realizations of the stochastic noise. 

In contrast to the 1D-ASEP particle conservation in the bulk does not imply 
that the current is the same for every link. 
Therefore, the current can be defined in several ways. An obvious choice 
is to count the number of particles entering or leaving the system 
since the particle reservoirs are each connected via a single link.
The incoming current $J_{in}$
through the entrance and the outgoing current $J_{out}$ from the exit 
are given by
\begin{eqnarray}
J_{in} &=& \alpha ( 1 - \langle n_1\rangle) \ , \label{J_in}\\
J_{out} &=& \beta \langle n_L \rangle \ . \label{J_out}
\end{eqnarray}
In the bulk site $i$~($=1,\cdots,L-1$), the current $J_i$ is 
defined as the total current of particles departing from sites 
$j=1,\cdots,i$ and arriving at sites $l=i+1,\cdots,L$. It is given by
\begin{equation}
J_i = \sum_{j=1}^i \sum_{l=i+1}^L u_{lj} \langle n_j (1-n_l)\rangle \ .
\end{equation}
With this definition of the current, the time evolution of the density is
given by $d\langle n_i\rangle/dt = J_{i-1}-J_i$ for $1<i<L$ and  
$d\langle n_1 \rangle/dt = J_{in} - J_1$ and 
$d\langle n_L\rangle/dt = J_{L-1} - J_{out}$.
Hence, all values of the local currents $J_i$, $J_{in}$ and $J_{out}$ 
should be the same in the steady state, where $d\langle n_i\rangle /dt = 0$.

The time-evolution equations (\ref{dni_dt}), (\ref{dn1_dt}), and 
(\ref{dnL_dt}) already illustrate the difficulty in calculating 
the quantities of  interest. 
In order to determine the particle density two-point correlations have 
to be calculated,  which in turn require higher order correlation functions. 
Furthermore, the quenched average over the random variables
$\{u_{ij}\}$ is necessary. 
An exact solution of this process is not available.  So, we will investigate the model 
using a numerical simulation method in the following section. 

\section{Simulation Results}\label{sec3}
In this section we present numerical results obtained from Monte Carlo 
simulations. For a simulation, one generates a graph $\mathcal{G}$ with an
adjacency matrix $\bf{A}$
consisting of a 1D chain of $L$ sites and $pL$ shortcuts. 
Particles on $\mathcal{G}$ move in the following way: First, we select a random 
variable $l\in \{0,1,\cdots,L\}$ with equal probability. (i) If
$l=0$, we add a particle to the entrance site $i=1$ with 
probability $\alpha$ if it is empty. (ii) If $0<l<L$, we try to move a particle (if any) 
at site $l$ to a target site $j$ selected among $\{l+1,\cdots,L\}$ with 
probability $u_{jl}$ given in (\ref{p_hop}). The particle move can be carried out if the target site is empty. (iii) If $l=L$, we remove a
particle (if any) at the exit site $i=L$ with probability $\beta$. 
The time is incremented by unity after $(L+1)$ trials. 

We are interested in the particle distribution and the current in the steady
state. After a transient period of time interval $T_t$, we average the 
occupation number for the time interval $T_s$ to obtain
the steady-state occupation number distribution $\langle n_i\rangle$. 
Note that the mean occupation number varies from one 
realization of a graph to another. 
In a second step a quenched average $[(\cdots)]_{\mathcal{G}}$ over graph realizations 
is necessary. The quenched averaged quantities will be denoted as 
\begin{equation}
\rho_i = [\langle n_i\rangle]_{\mathcal{G}} \ .
\end{equation}
The overall particle density is given by
\begin{equation}
\rho = \frac{1}{L}\sum_{i=1}^L \rho_i \ .
\end{equation}
The steady-state current is given by
\begin{equation}
J = \alpha (1-\rho_1) = \beta \rho_L \ .
\end{equation}

\begin{figure}[t]
\includegraphics*[width=\columnwidth]{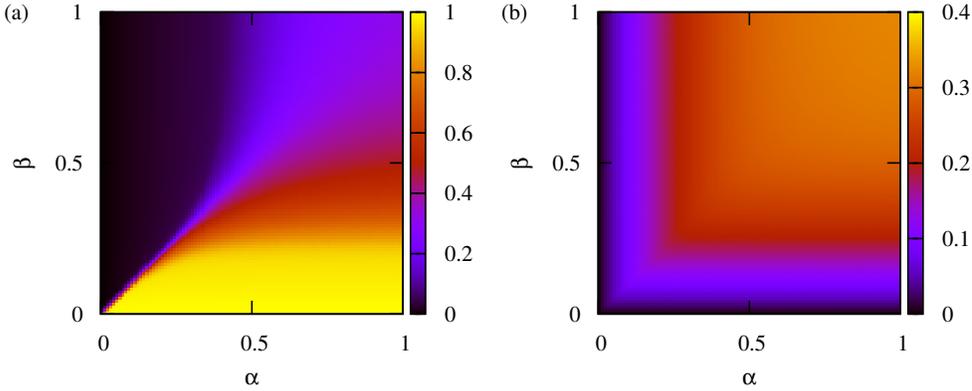}
\caption{(Color online) Particle density $\rho$ (a) and current $J$
(b) in the $\alpha\beta$ plane. The system size is $L=400$ and 
the shortcut density $p=0.2$.}\label{fig2}
\end{figure}

The overall behavior of the particle density $\rho$ and the current $J$
is presented in Figure~\ref{fig2}.
Those data were averaged over
$N_s = 1000$ graph realizations 
over a time interval $T_s = 500000$ 
after a transient interval $T_t = 10000$.
Figure~\ref{fig2} suggests that there exist three different regimes. When
$\alpha$ is small, the overall density $\rho$ is close to zero.
On the other hand, $\rho$ is close to 1 when $\beta$ is small. 
A finite-size-scaling~(FSS) analysis
reveals that the system indeed undergoes nonequilibrium phase transitions
between three different phases, which will be discussed below.

\begin{figure}[t]
\includegraphics*[width=\columnwidth]{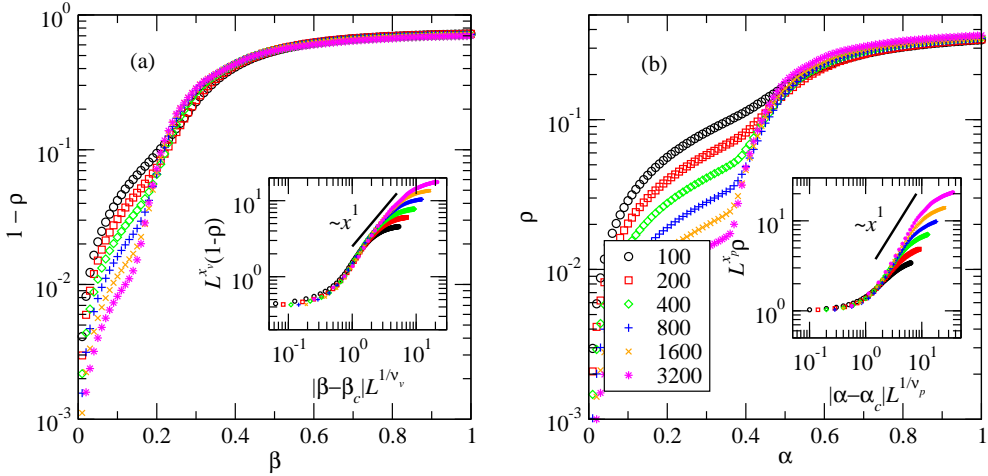}
\caption{(Color online) (a) Plot of the density of empty sites vs. $\beta$ 
at fixed $\alpha=0.8$. (b) Plot of the particle density vs. $\alpha$ at
fixed $\beta=0.8$.}\label{fig3}
\end{figure}

Figure~\ref{fig3}(a) shows the plot of $(1-\rho)$ along a line 
$\alpha=0.8$ at several values of $L=100,\cdots,3200$. Those data
are well fitted to a FSS form~\cite{Privman}
\begin{equation}
(1-\rho) = L^{-x_v} \mathcal{F}_v((\beta-\beta_c)L^{1/\nu_v})
\end{equation}
with $\beta_c \simeq 0.15$, $x_v \simeq 0.4$, and $\nu_v \simeq 2.5$.
The scaling function has a limiting behavior 
$\mathcal{F}_v(y \gg 1) \sim y^{\beta_v'}$ with 
$\beta_v' = x_v \nu_v \simeq 1.0$. 
The FSS indicates that system undergoes a
continuous phase transition at a critical point $\beta = \beta_c$.
For $\beta<\beta_c$, the density of vacant sites is zero. That is to say,
the system is fully occupied by particles. Such a macroscopic state
will be called a {\em jammed phase}. 
Near the critical point with $\epsilon = \beta-\beta_c$, 
the density of empty sites scales as
\begin{equation}
(1-\rho) \sim \epsilon^{\beta_v'} \ .
\end{equation}

Figure~\ref{fig3}(b) shows the plot of $\rho$ along a line 
$\beta=0.8$. The data are well fitted to a FSS form~\cite{Privman}
\begin{equation}
\rho = L^{-x_p} \mathcal{F}_p((\alpha-\alpha_c)L^{1/\nu_p})
\end{equation}
with $\alpha_c \simeq 0.36$, $x_p \simeq 0.5$, and $\nu_p \simeq 2$. The
scaling function has a limiting behavior $\mathcal{F}_p(y\gg 1)\sim
y^{\beta_p'}$ with $\beta_p' = x_p \nu_p \simeq 1.0$. Hence we conclude that
the system undergoes a continuous phase transition at a
critical point $\alpha=\alpha_c$. The particle density vanishes for
$\alpha<\alpha_c$. Such a macroscopic state will be called 
an {\em empty phase}.
Near the critical point with $\epsilon =
\alpha-\alpha_c$, the particle
density scales as 
\begin{equation}
\rho \sim \epsilon^{\beta_p'} \ .
\end{equation}

\begin{figure}[t]
\includegraphics*[width=0.6\columnwidth]{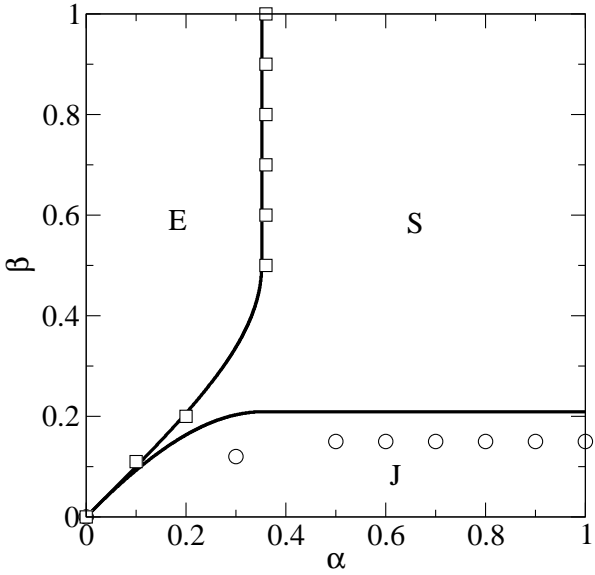}
\caption{Phase diagram at $p=0.2$. The location of symbols~($\Box$
between the E and S phases, and $\circ$ between the S and J phases) is
determined from the FSS analysis. The solid lines are the phase boundary
obtained from the mean-field and annealed approximation developed in
Sec.~\ref{sec4}.}
\label{fig4}
\end{figure}
Repeating the FSS analysis, we obtain the numerical phase diagram as
shown in Figure~\ref{fig4}. The phase diagram consists of the empty~(E) phase
with $\rho=0$ and the jammed~(J) phase with $\rho=1$. The other phase 
with $0<\rho<1$ will be called a shock~(S) phase.
Although the phase diagram looks similar to that of the ASEP on 1D 
chains~\cite{Derrida93,Schutz93},
the nature of the phases is different.

\begin{figure}[t]
\includegraphics*[width=\columnwidth]{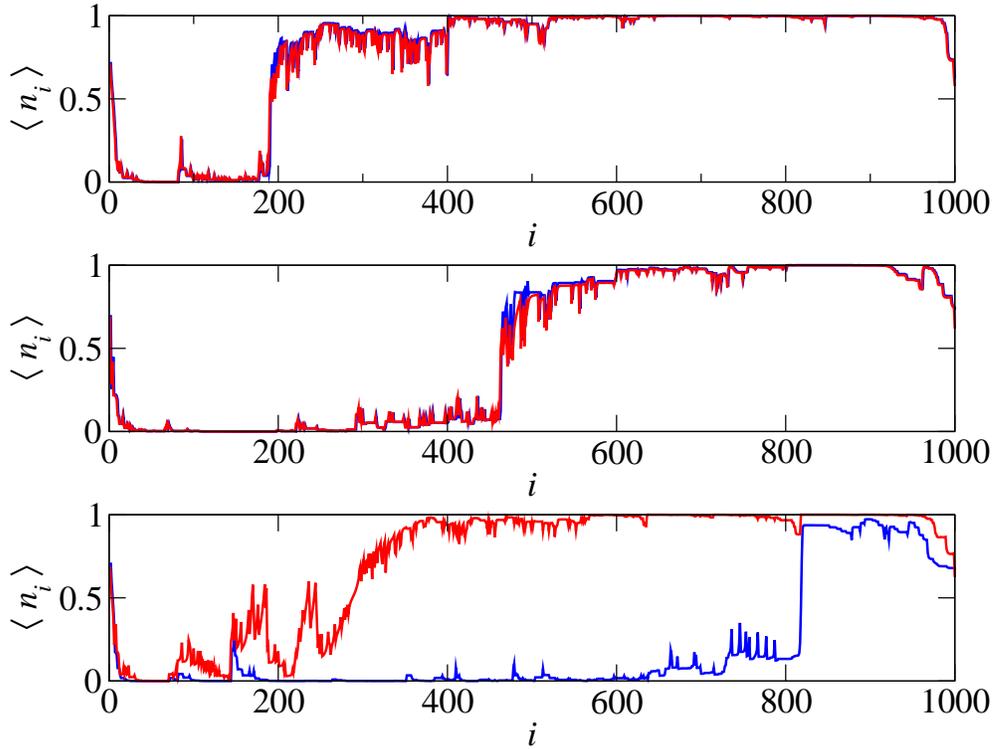}
\caption{(Color online) Particle density profile in the shock phase. 
Each panel contains a data set obtained from a different realization of a
graph. Monte Carlo simulation data are drawn with a red line. 
Also shown with a blue line are numerical data from the mean field 
approximation as explained in Sec.~\ref{sec4}.
Parameter values are $L=1000$, $p=0.2$, $\alpha=1.0$, and $\beta=0.5$.}
\label{fig5}
\end{figure}

The particle distribution is intriguing in the S phase.
Taking $\alpha=1.0$ and $\beta=0.5$, we have measured the
particle density distribution $\{\langle n_i\rangle\}$ in the steady state
at a given realization of $\mathcal{G}$.
Figure~\ref{fig5} shows typical distributions.
There is a phase separation between a region with $\langle
n_i \rangle \simeq 0$ and a region with $\langle n_i
\rangle \simeq 1.0$. The domain boundary between the two regions is called a
shock. 
The localized shock gives a hint why there are phase transitions into the E and J phases. 
The system can be in the E~(J) phase when the shock is absorbed at the
exit~(entrance). Hence it is crucial to understand the mechanism for the
shock formation.
This will be discussed in the following section.

Before proceeding, we present a physical argument for the phase diagram.
It is known that the diameter $l_D$ of the small-world network with nonzero
shortcut density $p$ scales as $l_D \sim \frac{1}{p} \ln (pL)$~\cite{BA}. 
This means that any site can be reached from a given site within 
$l_D$ steps~\cite{Watts98,BA}. 
Then, a particle injected
at the entrance can move toward the exit with a diverging speed 
$v \simeq L/l_D \sim L/\ln L$ along the backbone in the $L\to\infty$ limit 
when $\alpha$ is so small that the exclusion is irrelevant near 
the entrance. Consequently, particles can escape instantly from the system 
and the E phase can be realized. Using the same argument we can understand 
the origin of the J phase. A hopping of a particle from one site to another
is equivalent to a hopping of a hole in the opposite direction. 
Hence, when $\beta$ is small enough, a hole injected at the exit can travel 
toward the entrance with the diverging speed in $\mathcal{O}(\ln L)$ steps
and the J phase can be
realized. The S phase is a result of competition between the
empty domain~(which is stabilized by the long-range hopping of particles) 
and the jammed domain~(which is stabilized by the long-range hopping of
holes). This argument clearly shows that the small-world property 
is crucial in the formation of the E, J, and S phases. 
The small-world property emerges at any nonzero value of $p$~\cite{BA}. 
Hence we expect that those phases replace the LD, HD, and MC phases of the 
conventional ASEP in 1D chains immediately as one turns on the long-range
links.

We have also performed the numerical simulations with $p=0.1$ and $p=1.0$. 
In both cases we observe qualitatively the same phase diagram.
As $p$ increases,
the E phase expands while the J phase shrinks. 
For example, when $\beta=0.8$, we found that 
$\alpha_c = 0.35$, $0.36$, and $0.39$ at $p=0.1$, $0.2$, and $1.0$,
respectively. When $\alpha = 0.8$, we found that 
$\beta_c = 0.20$, $0.15$, and $0.07$ at $p=0.1$, $0.2$,
and $1.0$, respectively. We do not find a simple explanation for the
observed $p$ dependence of the phase boundary.

\section{Analytic Results}\label{sec4}

\subsection{Mean field approximation}
We adopt a mean field approximation by assuming that 
\begin{equation}
\langle n_i n_j \rangle = \langle n_i \rangle \langle n_j \rangle
\end{equation}
for all $i$ and $j$. Applying the mean field approximation to 
(\ref{dni_dt}), (\ref{dn1_dt}), and (\ref{dnL_dt}) yields that 
\begin{equation}\label{dndt}
\frac{d}{dt}\langle n_{i}\rangle = -
\langle n_i \rangle R_i  + (1-\langle n_i \rangle) Q_i \ ,
\end{equation}
where the auxiliary quantities are defined as
\begin{equation}
R_i = \beta \delta_{i,L} +  \left(\sum_{j>i}u_{ji} 
\left(1-\langle n_j \rangle\right)\right)(1-\delta_{i,L})
\end{equation}
and 
\begin{equation}
Q_i \equiv \alpha \delta_{i,1} +  \left( 
\sum_{j<i} u_{ij} \langle n_j \rangle \right)(1-\delta_{i,1}) 
\end{equation}
with the Kronecker $\delta$ symbol.
The quantities $R_i$ and $Q_i$ can be interpreted as particle evaporation 
and deposition rates at site $i$, respectively. They depend on the
structure of an underlying graph through $\{u_{ij}\}$ 
and the whole particle density distribution $\{\langle n_i \rangle \}$.
This interpretation of the rates  $R_i$ and $Q_i$ allows to relate our model  
to the 1D ASEP with constant evaporation and deposition rates 
which was studied in \cite{Parmeggiani03,Evans03,Juhasz04}. 
In contrast to that model the rates  $R_i$ and $Q_i$ in our model 
depend on the particle distribution as well as on the realization 
of the adjacency matrix.

The steady-state density satisfying $d\langle n_i\rangle/dt=0$ is given by
\begin{equation}\label{SCE}
\langle n_i \rangle = \frac{Q_i}{R_i + Q_i} \ .
\end{equation}
Since $R_i$ and $Q_i$ depends on the particle distribution, 
(\ref{SCE}) should be solved self-consistently.
The self-consistent equation can be solved numerically via iteration.
Starting from any trial distribution, one updates it by
evaluating the right-hand side of (\ref{SCE}). 
A particle distribution converges to a steady-state distribution 
without difficulty.

In order to test the mean field approximation, we compare
the particle density profile obtained from the Monte Carlo method and
the mean field approximation. 
For a given graph $\mathcal{G}$, we have performed the Monte Carlo simulation 
and solved the self-consistent equation for
the steady-state density profile.
They are compared in Figure~\ref{fig5}. 
The mean field approximation reproduces the phase separation and the shock 
in the density profile. 
In many cases, a mean field result is remarkably close to
a Monte Carlo result~(see the top and middle
panels in Figure~\ref{fig5}). On the other hand, in some cases, there is a
noticeable quantitative discrepancy between them~(see the bottom panel in
Figure~\ref{fig5}). Nevertheless, the mean field result still indicates the 
presence of the shock clearly. 
Therefore, we conclude that the mean field theory is
suitable for the description of the phase transitions.

\subsection{Annealed network approximation}
The density profile given by the solution of
(\ref{SCE}) depends on a graph realization $\mathcal{G}$. 
We have to perform the quenched average over graph realizations 
to obtain the disorder-averaged density profile
$\rho_i = [\langle n_i\rangle]_{\mathcal{G}}$. 
The quenched average is analytically intractable. 
Hence we further make an approximation by replacing an adjacency matrix
element $A_{ij}$, which is a quenched random variable, 
with its disorder-averaged value $[A_{ij}]_{\mathcal{G}}$. Such an
approximation is called an annealed approximation. The annealed
approximation is useful in studying physical systems on graphs or
networks~\cite{Boguna09,Noh09,Lee09}. 

On a graph of $L$ sites, there are $pL$ long-range links. 
Hence the probability to find a long-range links between two sites 
$i$ and $j\neq i\pm 1$ is given by $p_1 = 2pL/((L-1)(L-2)) \simeq 2p/L$.
Short-range links are connecting sites $i$ and $i+1$ for $i=1,\cdots,L-1$.
Taking account of the short- and long-range links, we obtain that
\begin{equation}\label{A_av}
[A_{ij}]_{\mathcal{G}} = \delta_{i,j+1} + ( 1 - \delta_{i,j+1}) \frac{2p}{L}  
\end{equation}
for $i>j$ and $[A_{ij}]_{\mathcal{G}}=0$ for $i\le j$. 
The parameter
$u_{ij}$ is replaced by $u_{ij} = [A_{ij}]_{\mathcal G} / [k_j]_{\mathcal
G}$ with $[k_j]_{\mathcal G} = (1+2p(L-j-1)/L)$.

In the annealed approximation, the self-consistent
equation for the disorder-averaged density $\rho_i$ becomes 
\begin{equation}\label{rho_i}
\rho_i = \frac{Q_i}{R_i+Q_i}
\end{equation}
where $R_i$ and $Q_i$ are given by
\begin{eqnarray}
R_i &=& 1-\frac{1}{1+\frac{2p}{L}(L-i-1)} \left( \rho_{i+1} + \frac{2p}{L}
\sum_{j=i+2}^L \rho_j\right) \label{R_i_av} \\
Q_i &=& \frac{\rho_{i-1}}{1+\frac{2p}{L}(L-i)} + \sum_{j=1}^{i-2} 
\frac{ \frac{2p}{L} \rho_j }{1 + \frac{2p}{L} (L-j-1)} \label{Q_i_av}
\end{eqnarray}
with the boundary terms $R_L = \beta$ and $Q_1 = \alpha$.
Equations~(\ref{rho_i}), (\ref{R_i_av}), and (\ref{Q_i_av}) are the starting
point for further analysis.

\subsection{Shock state}
We first consider the sites with $i=\mathcal{O}(1)$ near the entrance. 
Ignoring $\mathcal{O}(L^{-1})$ corrections,
one can approximate $R_i$ and $Q_i$ as
\begin{eqnarray}
R_i^{in} & = & \frac{1}{1+2p} ( 1-\rho_{i+1}) + \frac{2p}{1+2p} (1-\rho) 
\label{R_in} \\
Q_i^{in} & = & \frac{1}{1+2p} \rho_{i-1}  \label{Q_in} 
\end{eqnarray}
with the boundary term $Q_1^{in}=\alpha$ and the overall particle density
$\rho$.
These expressions allow for an interpretation for an effective dynamics 
near the entrance: A particle at site $i$ performs a short-range jump 
to site $i+1$ with probability 
\begin{equation}
W^{in}_{h} = \frac{1}{1+2p}
\end{equation}
or annihilates spontaneously with the probability
\begin{equation}
W^{in}_{a}(\rho) = \frac{2p}{1+2p}(1-\rho) \ .
\end{equation}

Due to the effective annihilation, the particle density should
decay to zero exponentially with the distance from the entrance 
along the backbone with a characteristic 
length scale $\xi_{in} = 1/W_{a}^{in}(\rho)$ unless $\rho=1$.
This feature is consistent with the density profile shown in
Figure~\ref{fig5}. 

We next consider the sites $i=L-l$ with $l=\mathcal{O}(1)$ near the exit.
Ignoring again $\mathcal{O}(L^{-1})$ corrections, one can approximate
$R_i$ and $Q_i$ by
\begin{eqnarray}
R_i^{out} & = & 1-\rho_{i+1} \label{R_out}\\
Q_i^{out} & = & \rho_{i-1} +  \frac{2p}{L} 
          \sum_{j=1}^{i-2} \frac{\rho_j}{1+2p(L-j-1)/L}  
\label{Q_out}
\end{eqnarray}
with the boundary term $R_L^{out} = \beta$. 
These expressions suggest
that particles near the exit have a following effective dynamics:
A particle at site $i$ performs a short-range hopping to site $i+1$
with the probability
\begin{equation}
W^{out}_{h} = 1  \ ,
\end{equation}
and particles are created spontaneously at each site with the probability
given by the second term in (\ref{Q_out}).
Due to the creation, the particle density should saturate to unity as one
departs from the exit along the backbone.

The density profiles stemming from the both boundaries converge to different
values of 0 and 1. So, there must emerge a shock as a domain boundary. 
The position of the shock along the backbone 
is denoted by $i_S$, which is related to the overall particle density as 
\begin{equation}
\rho = 1 - \frac{i_S}{L} \ .
\end{equation}
Making use of the shock structure, the quantity $Q^{out}_i$ in
(\ref{Q_out}) is given by 
\begin{eqnarray}
Q^{out}_i &=& \rho_{i-1} + \frac{2p}{L} \sum_{j=i_S}^{i-2} 
 \frac{1}{1+2p(L-j-1)/L} \nonumber \\
&=& \rho_{i-1} + \ln(1+2p\rho)
\end{eqnarray}
with $\mathcal{O}(L^{-1})$ corrections being ignored. 
Correspondingly, the particle
creation probability near the exit is given by
\begin{equation}
W_c^{out}(\rho) = \ln(1+2p\rho) \ .
\end{equation}

So far we have established the shock state. Effectively, particles near the 
entrance hop to the right with the probability $W_h^{in}$ or are annihilated
with the probability $W_a^{in}(\rho)$. This dynamics results in a density
profile $\{\rho_i^{in}\}$ which decays to zero as $i$ increases. Near the
exit, particles are created effectively with the probability
$W_{c}^{out}(\rho)$ 
and hop to the right with the probability $W_h^{out}$.
This results in a density profile $\{\rho_i^{out}\}$ which converges to
unity as $i$ decreases from $L$. Both profiles should be matched at a
position $i_S=(1-\rho)L$ to yield a shock.

This situation is similar to the driven 1D ASEP with particle creation and 
annihilation~\cite{Parmeggiani03,Evans03,Juhasz04} or 
boundary driven multi-lane systems~(see e.g. 
\cite{reichenbach_f_f06, Schiffmann_10}). When the creation and
annihilation rates are spatially uniform and inversely proportional to the
lattice size, the system also develops a shock in the stead-state 
density profile~\cite{Parmeggiani03,Evans03,Juhasz04}. 
In comparison with the model studied 
in~\cite{Parmeggiani03,Evans03,Juhasz04,reichenbach_f_f06, Schiffmann_10}, 
the creation and annihilation rates are
not uniform in space: Particles are annihilated near the
entrance and created near the exit. The difference results in the feature
that the shock separates the empty and the fully jammed domains.

The parameters $W_a^{in}(\rho)$ and $W_c^{out}(\rho)$ depend on the
the overall particle density $\rho = 1-i_S/L$, which should be determined
self-consistently. The overall density can be obtained from the 
current conservation, which will be explained in the following subsection.

\subsection{Phase diagram}

Particles are injected at the entrance ($i=1$), move to the right, and are
removed at the exit~($i=L$). So, the system can carry a nonzero current. 
The incoming current at the entrance and outgoing current at the exit 
are given by
\begin{eqnarray}
J_{in}(\rho) &=& \alpha (1-\rho_1^{in}) \label{Jin} \\
J_{out}(\rho) &=& \beta \rho_L^{out} \ . \label{Jout}
\end{eqnarray}
Because $\rho_1^{in}$ and $\rho_L^{out}$ are governed by the
$\rho$-dependent effective dynamics, the incoming and the outgoing currents 
are given as a function of $\rho$.
Particle number conservation requires that the incoming and outgoing currents
should be the same in the steady state. 
The equality $J^{in}(\rho) = J^{out}(\rho)$ determines the overall particle
density $\rho$, hence the phase diagram.

The effective dynamics is still too complex and does not allow for the
closed-form solution for $\rho_i^{in,out}(\rho)$. Therefore, the exact phase diagram will be obtained from numerical solutions of the self-consistent equations and the
current-balance condition. 
Before doing so, we apply an approximate scheme to the self-consistent equations
in order to gain a physical insight.

In terms of the effective dynamics, all particles introduced at the entrance 
have to be transferred toward the exit and therefore annihilated from the entry area. So, the incoming current can also be written as
\begin{equation}\label{Jin_sum}
J_{in}(\rho) = W_a^{in} \sum_{i\geq 1} \rho_i^{in} \ .
\end{equation}
Using the explicit forms given in (\ref{R_in}) and (\ref{Q_in}),
the self-consistent equation for the density becomes as
\begin{equation}
\rho_i^{in} = \frac{\rho_{i-1}^{in}}{ 1+2p(1-\rho) + (\rho_{i-1}^{in} -
\rho_{i+1}^{in})} 
\end{equation}
for $i\geq 2$. Because of the 
continuous annihilation of particles, 
we expect that $\rho_{i}^{in}$ decays monotonically and rather fast to zero. 
In order to gain a qualitative understanding we can ignore 
$(\rho_{i-1}^{in} - \rho_{i+1}^{in})$ in the denominator to obtain that
\begin{equation}
\rho_{i}^{in} \simeq \frac{\rho_1^{in}}{ (1+2p(1-\rho))^{i-1} } \ .
\end{equation}
Inserting these approximate solutions into (\ref{Jin_sum}), one obtains
that 
\begin{equation}\label{Jin_trial}
J_{in} \simeq \frac{1+2p(1-\rho)}{1+2p}\rho_1^{in}.
\end{equation}
Comparing the two expressions for $J_{in}$ given in (\ref{Jin}) and
(\ref{Jin_trial}), one finds a solution for $\rho_1^{in}(\rho)$, which
yields that 
\begin{equation}\label{Jin_app}
J_{in}(\rho) \simeq \frac{\alpha(1+2p -2p \rho)}
{(1+\alpha)(1+2p) - 2p \rho} \ .
\end{equation}
It is a decreasing function of $\rho$ with
$J_{in}(0) = \frac{\alpha}{1+\alpha}$ and $J_{in}(1) =
\frac{\alpha}{1+\alpha(1+2p)}$.

One can carry out a similar analysis to obtain an approximate expression for
$J_{out}(\rho)$. First, the outgoing current given in (\ref{Jout})
should be equal to the total particle creation rate, that is to say,
\begin{equation}
J_{out} = W_c^{out} \sum_{i\leq L} (1 -\rho_{i}^{out}) \ .
\end{equation}
It is easy to show that the void density $(1-\rho_i^{out})$ 
satisfies self-consistent equations
\begin{equation}
1-\rho_{i}^{out} = \frac{1-\rho_{i+1}^{out}}{1+\ln(1+2p\rho) +
(\rho_{i-1}^{out}-\rho_{i+1}^{out})} 
\end{equation}
for $i<L$. Again the transport of particles via long ranged links can be understood as  spontaneous creation of particles at sites close to the exit. Therefore, we expect that
$(1-\rho_i^{out})$ decays to zero as $i$ decreases from $L$.
So we can ignore $(\rho_{i-1}^{out}-\rho_{i+1}^{out})$ in the
denominator as in the previous case. A similar calculation then yields 
\begin{equation}\label{Jout_app}
J_{out}(\rho) \simeq 
\frac{\beta(1+\ln(1+2p\rho))}{1+\beta+\ln(1+2p\rho)} \ .
\end{equation}
This is an increasing function of $\rho$ with
$J_{out}(0) = \frac{\beta}{1+\beta}$ and $J_{out}(1) =
\frac{\beta(1+\ln(1+2p))}{1+\beta+\ln(1+2p)}$.

\begin{figure}[t]
\includegraphics*[width=\columnwidth]{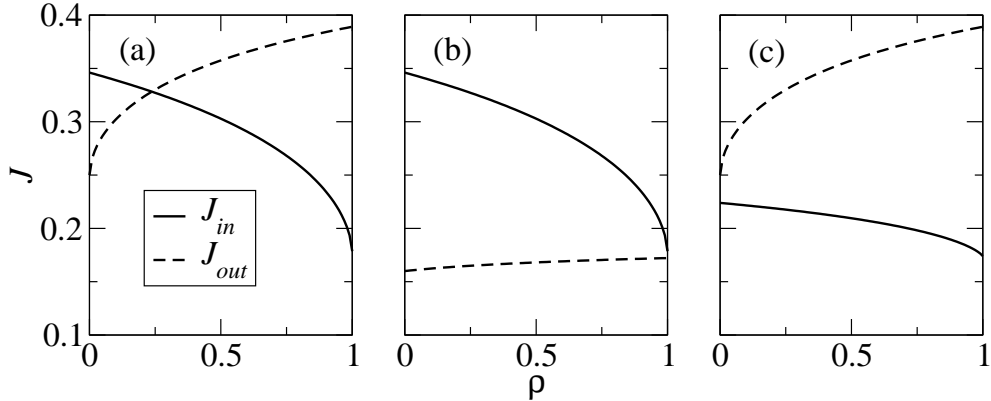}
\caption{Schematic plots of $J_{in}$~(solid lines) and $J_{out}$~(dashed
lines) against $\rho$.}\label{fig6}
\end{figure}

The particle number conservation requires that the incoming current and 
the outgoing current should be the same in the steady state. 
Figure~\ref{fig6} shows schematic plots of
$J^{in}(\rho)$ and $J^{out}(\rho)$ in three different situations. 
In the first case, the current curves may intersect with each other
at $\rho=\rho_0$ with $0<\rho_0<1$ as shown in Figure~\ref{fig6}(a). 
This case corresponds to the shock phase. The
intersection determines the shock position $i_S = (1-\rho_0)L$. 
If $\rho$ becomes greater~(smaller) than $\rho_0$ due 
a stochastic fluctuation, then the incoming current becomes
smaller~(greater) than the outgoing current. 
Consequently, the particle density is attracted toward the steady-state
value and the shock is driven toward the steady-state position. 
This explains the reason why there is a localized sharp
shock~\cite{Parmeggiani03,Evans03,Juhasz04}.

The second case with $J_{in}(\rho) > J_{out}(\rho)$ for all values of 
$\rho$ is shown in Figure~\ref{fig6}(b). Then, the
shock is localized at the entrance and the steady-state density is 
equal to one. In this case the system belongs to the jammed phase.

The third case is sketched in Figure~\ref{fig6}(c). If $J_{in}(\rho) <
J_{out}(\rho)$ for all values of $\rho$, the shock is localized at the exit
and the steady-state density is equal to zero. Then the system belongs to the
empty phase.

We have solved numerically exactly the self-consistent equations for 
$\rho^{in}_i$ and $\rho^{out}_i$ to obtain the incoming and outgoing 
currents as a
function of $\rho$ at each set of values of $\alpha$, $\beta$, and $p$.
The current balance condition allows us to evaluate the steady-state 
particle density and the current, hence the phase diagram. The resulting
numerical phase diagram in the $\alpha\beta$ plane is presented in 
Figure~\ref{fig4}. 
The phase diagram at $p=0.2$ consists of the empty phase, jammed
phase, and the shock phase, which is consistent with the Monte Carlo result. 
There is a quantitative discrepancy in the location of the phase boundaries, which is caused
by the approximations. 

The annealed approximation requires the self-averaging
property~\cite{Wiseman98,Roy06}. The self-averaging property in the small-world
network was tested for the equilibrium Ising model~\cite{Roy06}.
To test the self-averaging property in our nonequilibrium model, we
have measured the relative sample-to-sample fluctuation of the particle
density $X\equiv \sqrt{ \left(
[\langle \frac{1}{L}\sum_i n_i\rangle^2]_{\mathcal G} 
- [\langle \frac{1}{L} \sum_i n_i \rangle]_{\mathcal G}^2 \right)} / 
[\langle \frac{1}{L} \sum_i n_i \rangle]_{\mathcal G}$. In the E and J
phases, it decays to zero as $X \sim L^{-1/2}$ suggesting a strong
self-averaging~\cite{Wiseman98}. On the other hand, it converges to a finite
value in the S phase, an indication of non-self-averaging. This tells us
that the annealed approximation has a limitation. It explains successfully
the mechanism of shock formation, but not its average position due to 
the strong fluctuations. 
A refined approach beyond the annealed approximation is necessary to study 
the sample-to-sample fluctuation phenomena in the S phase.

\section{Summary and Conclusion}\label{sec5}
We have studied the boundary-driven ASEP in the 1D chain with long-range 
links. 
This setup represents a generic situation for directed transport in a complex 
network, e.g., the exchange of data between two sites of a computer network. 

The backbone of nearest neighbor links ensures the existence of a path 
between start and destination.  The long ranged links add shortcuts 
to the transport network that are in principle able to enhance 
the capacity of the system.

Considering the typical results for transport problems on complex networks 
one would also expect that long-range shortcuts,
added randomly on to a lattice, are believed to suppress fluctuations and
make a system homogeneous. 
This is, however, not the case when there is a boundary driving. 
Compared to the pure one-dimensional system it turns out that 
the long-range links play an essential role. 
They generate a localized shock which separates the 1D chain 
backbone into an empty and jammed regions. 

Adopting the mean field and the annealed approximations, 
we have derived effective dynamics near the entrance and the exit, 
which are similar to those of the ASEP with 
spontaneous particle annihilation and creation, respectively. 
The effective theory reveals the mechanism for the shock formation and for
the phase transition. The phase diagram consists of the empty, shock, and 
jammed phases. 
The shock phase is characterized by presence of a  localized shock 
and separating a low and a high density domain. 
The shock position and the overall particle
density $\rho$ vary continuously with the model-parameters. In
the empty~(jammed) phase, the shock is anchored to the exit~(entrance) to 
yield $\rho=0~(1)$.

In conclusion, our study shows that a driven system on a spatially 
disordered structure displays an inhomogeneous pattern. 
The appearance of a localized shock is reminiscent of one-dimensional 
systems without particle-conservation in the bulk. 
In contrast to these systems the mechanism driving the  
localization of the  shock is neither the competition between bulk 
and boundary reservoirs nor an optimal partitioning between multiple lanes. 
Here, the long-ranged links enhance the mass transfer between entry and 
exit and thereby stabilize the position of the shock. 
   
Considering more generally the transport capacity of a complex network 
between arbitrary sites our results have important consequences. 
They indicate that, as far as the capacity of the feeding particle 
reservoir does not exceed the capacity of the exit reservoir, 
the sites of the backbone are only rarely occupied and can be used 
in parallel for transport issues between other sites. Contrary, 
overfeeding the backbone leads to a complete blockage of the 
sites in question which may spread over the whole network.    

\ack This work was supported by Mid-career Researcher Program through NRF
grant (No.~2010-0013903) funded by the MEST.

\end{document}